\documentclass[letter,longauth,traditabstract]{aa}
\usepackage{graphicx}
\usepackage{txfonts}
\usepackage{natbib}
\usepackage{color}
\bibpunct{(}{)}{;}{a}{}{,} 

\begin{document}
\title{\rm Detection of warm water vapour in Taurus protoplanetary
  discs by Herschel} \subtitle{}
\author{P. Riviere-Marichalar\inst{1}, F. M\'enard\inst{2},
  W. F. Thi\inst{2}, I. Kamp\inst{3}, B. Montesinos\inst{1},
  G. Meeus\inst{4}, P. Woitke\inst{5,6,7}, C. Howard\inst{8},
  G. Sandell\inst{8}, L. Podio\inst{3}, W. R. F. Dent\inst{9},
  I. Mendigut\'{i}a\inst{1}, C. Pinte\inst{2}, G.J. White\inst{10,11},
  D. Barrado\inst{1,12} }

\institute{Centro de Astrobiolog\'{\i}a -- Depto. Astrof\'{\i}sica
  (CSIC--INTA), POB 78, 28691 Villanueva de la Ca\~nada, Spain\\ 
  \email{riviere@cab.inta-csic.es} \and UJF-Grenoble 1 / CNRS-INSU,
  Institut de Plan\'{e}tologie et d'Astrophysique (IPAG) UMR 5274,
  Grenoble, F-38041, France 
  \and Kapteyn Astronomical Institute, P.O. Box 800, 9700 AV
  Groningen, The Netherlands 
  \and Dep. de F\'isica Te\'orica, Fac. de Ciencias, UAM Campus
  Cantoblanco, 28049 Madrid, Spain 
  \and University of Vienna, Dept. of Astronomy,
  T\"urkenschanzstr. 17, A-1180 Vienna, Austria 
  \and UK Astronomy Technology Centre, Royal Observatory, Edinburgh,
  Blackford Hill, Edinburgh EH9 3HJ, UK 
  \and SUPA, School of Physics \& Astronomy, University of
  St. Andrews, North Haugh, St. Andrews KY16 9SS, UK 
  \and SOFIA-USRA, NASA Ames Research Center 
  \and ALMA, Avda Apoquindo 3846, Piso 19, Edificio Alsacia, Las
  Condes, Santiago, Chile 
  \and Department of Physics \& Astronomy, The Open University, Milton
  Keynes MK7 6AA, UK
  \and The Rutherford Appleton Laboratory, Chilton, Didcot, OX11 OQL,
  UK 
  \and Calar Alto Observatory, Centro Astron\'{o}mico Hispano-Alem\'{a}n C/Jes\'{u}s Durb\'{a}n Rem\'{o}n, 2-2, 04004 Almer\'{i}a, Spain 
} \authorrunning{P. Riviere-Marichalar} \date{}

 \abstract{

Line spectra of 68 Taurus T Tauri stars were obtained
     with the \textit{Herschel-PACS (Photodetector Array Camera
  \& Spectrometer)} instrument as part of the
     \textit{GASPS (Gas Evolution in Protoplanetary Systems)} survey of protoplanetary discs. A careful
     examination of the linescans centred on the [OI] 63.18 $\rm \mu m$
     fine-structure line unveiled a line at 63.32 $\mu$m in some of these
     spectra.  We identify this line with the $\rm 8_{18} \to 7_{07}$
     transition of ortho-water. It is detected confidently
     (i.e., $\rm >3 \sigma$) in eight sources, i.e., $\sim24\%$ of the
     sub-sample with gas-rich discs. Several statistical tests were used to
     search for correlations with other disc and stellar
       parameters such as line fluxes of [\ion{O}{I}] 6300\AA~ and
     63.18~$\mu$m; X-ray luminosity and continuum levels at 63~$\mu$m and
     850~$\mu$m.  Correlations are found between the water line
     fluxes and the [\ion{O}{I}] 63.18 $\mu$m line luminosity, the
     dust continuum, and possibly with the stellar X-ray luminosity.  This is
     the first time that this line of warm water vapour has been detected in
     protoplanetary discs.  We discuss its origins, in particular
     whether it comes from the inner disc and/or disc surface or from
     shocks in outflows and jets. Our analysis favours a disc origin, with the observed water
     vapour line produced within 2-3AU from the central stars, where
     the gas temperature is of the order of 500-600K.}
\keywords{ Stars: formation, astrobiology, astrochemistry, line: identification, molecular data, protoplanetary discs.}
   \maketitle
\section{Introduction} 

Discs are natural by-products of star formation and the birthplaces of
planets. One of the key questions intimately linked with planet
formation and the concept of planet habitability is how much vapour and icy
water is present in discs and how it is radially distributed. However, it is only recently that
observations of water vapour in discs have become possible.

\cite{CarrNajita2008}, using the \textit{Spitzer} InfraRed
Spectrograph (IRS), reported a rich molecular emission-line spectrum
dominated by rotational transitions of hot water from the disc of \object{AA
  Tau}. They concluded that the molecular emission seen in the
mid-IR has a most likely origin within the 2-3 AU inner regions of the disc.
\cite{Salyk2008} detected water emission in the 10-20 $\rm \mu m$
region with Spitzer-IRS, as well as water and hydroxyl emission around
3 $\rm \mu m$ with \textit{NIRSPEC} on \textit{Keck II}, for both \object{DR Tau} and \object{AS 205 A}. The emission
comes from the disc atmospheres of the objects and the excitation temperatures were found to be ($\rm \sim 1000$ K), which is typical of terrestrial planet formation regions. \cite{Pontoppidan2010_0} performed a survey for more protoplanetary discs
in Ophiuchus, Lupus, and Chamaeleon, again with \textit{Spitzer-IRS}, and
concluded that the presence of mid-IR molecular emission lines,
including those of water, is a common phenomenon in discs around Sun-like
stars. Also, \cite{Pontoppidan2010} presented a sample of ground-based
observations of pure rotational lines of water vapour in the
protoplanetary discs of \object{AS 205 A} and \object{RNO 90} that was analysed to measure
 line widths of 30--60 km $\rm s^{-1}$, which is consistent with an
origin in a disc in Keplerian rotation at a radius of $\rm \sim 1$ AU,
and gas temperatures in the range 500--600 K. 

\begin{figure*}
  \centering
 \includegraphics[scale=0.9]{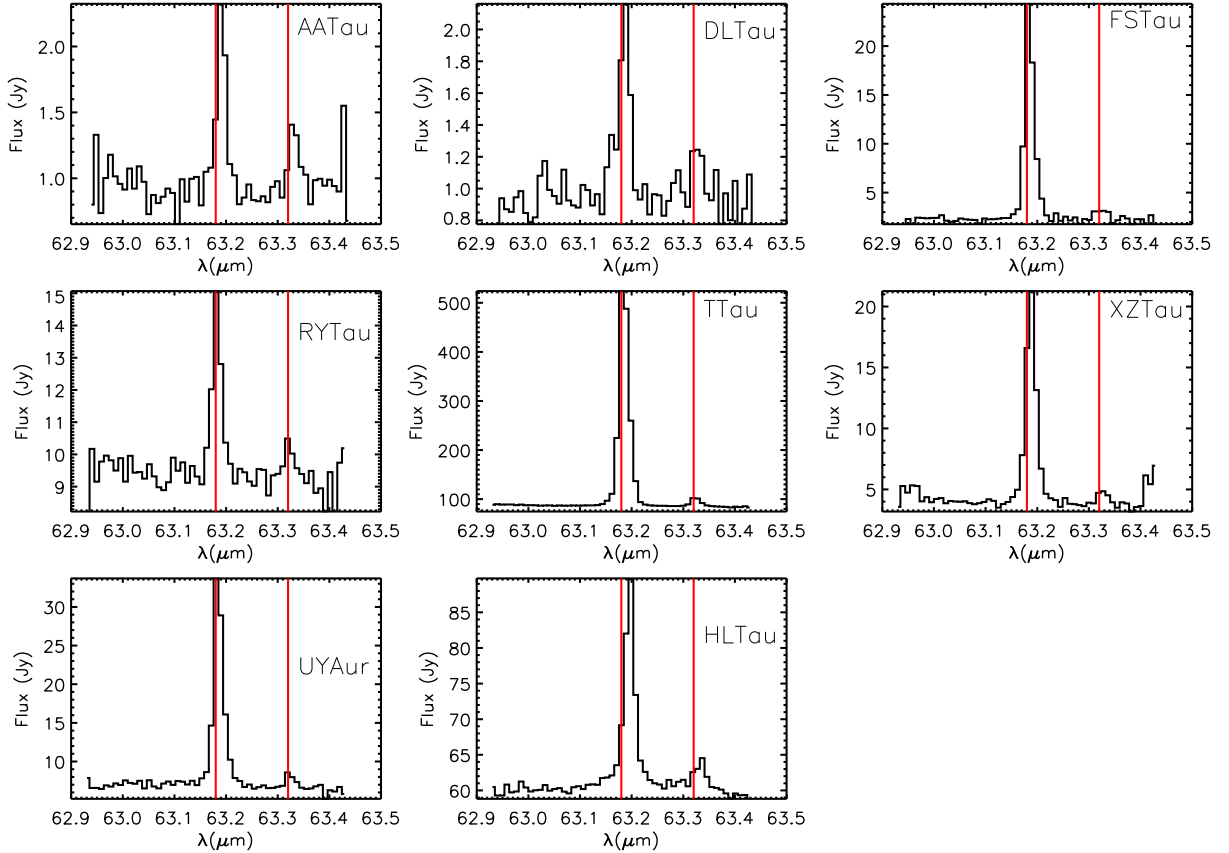}
  \caption{Spectra for the objects with a $\rm 63.32 \mu m$ feature detection ($\rm > 3 \sigma$). The red lines indicate the rest wavelength of the [\ion{O}{I}] and o-$\rm H_{2}O$ emission.}
  \label{detectionSpec}
\end{figure*}
The \textit{Herschel Space Observatory} \citep{Pilbratt2010} has
opened the far-IR window with unprecedented sensitivity, allowing astronomers to
survey the atomic and molecular gas content of disc regions that are cooler
than those probed by \textit{Spitzer} and ground-based
instruments. Predictions of the water lines detectable by Herschel can be found in \citet{Cernicharo2009} and \citet{Woitke2009}. \cite{Hogerheijde2011} presented the first detection of cold water in the disc of the $\rm \sim$10 Myr old TW Hya,
using \textit{Herschel-HIFI}.  The \textit{Herschel} Open Time Key Program \textit{GASPS (Gas Evolution in Protoplanetary Systems)}
\citep{Mathews2010} is conducting a survey to measure gas lines and continuum in $\sim$250 discs around low- and intermediate-mass stars with ages in the range 1-30Myr with PACS, the \textit{Photodetector Array Camera
  \& Spectrometer} \citep{Poglitsch2010}. 

In this letter, we report the first detection of the o-$\rm H_{2}O$
line at 63.32 $\rm \mu m$ in a subsample of protoplanetary discs
around T Tauri stars in the 1-3 Myr old Taurus star forming region.

\section{Observations and data reduction}
This study is based on a sample of 68 classical and weak-line T Tauri
stars from the Taurus star forming region with spectral
measurements from \textit{Herschel}-PACS centred at the wavelength of
the [\ion{O}{I}] $\rm ^{3}P_{1} \rightarrow$$\rm ^{3}P_{2}$ 63.184 $\rm
\mu m$ line. The Taurus star-forming region is one of the main targets
for the study of protoplanetary systems, because it is among the
nearest star-forming regions (d = 140 pc) with a well-known population of more than 300 young
stars and brown dwarfs according to \cite{Kenyon2008},
\cite{Luhman2010}, and \cite{Rebull2010}.

The observations described in this letter are part of the
\textit{Herschel} Open Time Key Programme \textit{GASPS}
[P.I. W. Dent],  \citep[see]{Mathews2010}, a flux-limited survey devoted
to the study of the gas and dust in circumstellar systems around young
stars. The survey focuses on the detection of the [\ion{O}{I}]
emission at the 63.18 $\rm \mu m$ feature. For this study, we analyzed 68 stars with spectral types ranging from late
  F-early G to mid M.  The PACS spectral observations were made in
chop/nod pointed line mode. The observing times ranged from 1215 to
6628 seconds, depending on the number of nod cycles. The data were
reduced using HIPE 7.0.1751. A modified version of the PACS pipeline was used, which included: saturated and bad pixel removal, chop
subtraction, relative spectral response-function correction, and flat
fielding. Many observations suffer from systematic pointing errors,
in some cases as large as 8$\rm \arcsec$, and are always
shifted to the East. This is due to a plate scale error in the star
tracker, which is normally negligible except in areas where the
tracked stars are asymmetrically distributed within the field, as in
Taurus. The mis-pointing translates into systematic small shifts in
the line centre position. When the star was well centred within a single spaxel, we extracted the flux from that spaxel 
and applied the proper aperture correction. When the flux was spread over more than one spaxel, we co-added the spaxels.

\begin{table}[th]
\caption{Line positions and fluxes from PACS spectra.}             
\label{fitTable}      
\centering          
\begin{tabular}{lllll}     
\hline\hline       
Name & Sp. type & $\rm \lambda _{H_{2}O} - \lambda _{[OI]}$ & [\ion{O}{I}] flux  & $\rm o-H_{2}O$ flux    \\ 
\hline
	 -- & -- & $\rm \mu m$&$10^{-17} $ W/ $\rm m^{2}$ & $10^{-17} $ W/$\rm m^{2}$ \\ 
\hline                    
  \object{AA Tau} & K7& 0.140 &  2.2 $\rm \pm$ 0.13  & 0.80 $\rm \pm$ 0.13\\
  \object{DL Tau} & K7  & 0.141 &  2.4 $\rm \pm$ 0.15 & 0.65 $\rm \pm$ 0.14\\
  \object{FS Tau} &  M0 & 0.139 & 37 $\rm \pm$ 0.26  & 2.00 $\rm \pm$ 0.33\\
  \object{RY Tau} & K1  & 0.139 &10 $\rm \pm$ 0.42 & 1.95 $\rm \pm$ 0.38 \\
  \object{T Tau} & K0  & 0.138 & 830 $\rm \pm$ 0.75 & 27.8 $\rm \pm$ 0.7 \\
  \object{XZ Tau} & M2 & 0.139 & 32.2 $\rm \pm$ 0.48 & 2.11 $\rm \pm$ 0.48 \\
  \object{HL Tau} & K7 & 0.142 & 54.3 $\rm \pm$ 0.71& 8.14 $\rm \pm$ 0.80 \\
  \object{UY Aur} & M0  & 0.139 & 33.6 $\rm \pm$ 0.37 & 1.90 $\rm \pm$ 0.32 \\
\hline                  
\end{tabular}
\tablefoot{
All spectral types from the compilation of \cite{Luhman2010}.}
\end{table}

\section{Results and discussion.}
Among the sample of 68 Taurus targets studied in this letter, 33 have discs that are rich in gas. These 33 all show the [\ion{O}{I}] $\rm^{3}P_{1} \rightarrow^{3}\!\!P_{2}$ line in emission at 63.18 $\rm \mu m$ (signal-to-noise ratio
$>$ 3, with values ranging from 3 to 375). In 8 of these 33 targets ($\rm \sim$ 24$\%$), an additional
fainter emission-line at 63.32 $\rm \mu m$ is detected (Fig.1). We computed 63.32 $\rm \mu m$ line fluxes by fitting a gaussian plus continuum curve to the spectrum using DIPSO
\footnote{http://star-www.rl.ac.uk/docs/sun50.htx/sun50.html}. To improve the line fitting, the noisier 
edges of the spectral range were removed (i.e., $\rm \lambda < 63.0$ and $\rm \lambda > 63.4$).
The results are listed in Table \ref{fitTable}, where we
report the peak position of the feature with respect to the observed
wavelength of the [\ion{O}{I}] 63.18 $\rm \mu m$ line.  According to
these fits, the peak of the feature is at  $\rm
\lambda_{0} = 63.32~\mu$m. The FWHM is  0.020 $\rm \mu m$, i. e., the
instrumental FWHM for an unresolved line. We identify the feature as
the ortho-$\rm H_{2}O$ $\rm 8_{18} \to 7_{07}$ transition at 63.324
$\rm \mu m$ ($\rm E_{Upper~Level}$=1070.7~K, Einstein
A=1.751~$s^{-1}$) since no other abundant species emit strongly at or close to the
observed wavelength of the feature. This water feature was observed by \citet{Herczeg2011} in 
the outflow of NGC~1333~IRAS~4B.
\begin{figure}
    \includegraphics[scale=0.50,trim = 3mm 4mm 2mm 3mm,clip]{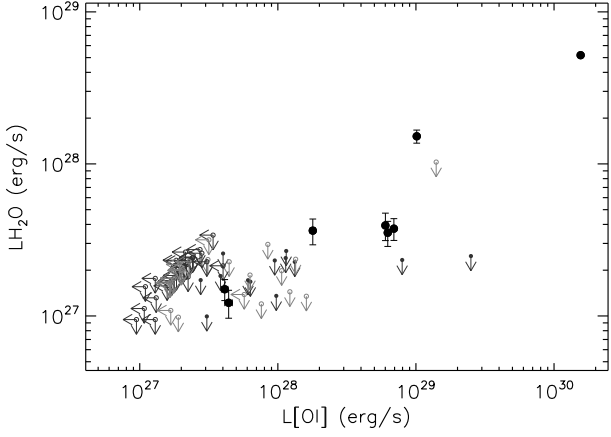}
  \caption{Plot of the 63.32 $\rm \mu$m o-H$_{2}\rm{O}$ line luminosity versus [OI] 63.18~$\rm \mu m$. Filled dots are detections, arrows are upper limits. Solid, dark grey arrows represent objects with non-detections spanning the same spectral range as the objects with detections. Light grey, empty arrows represent non-detections with other spectral types.}
  \label{H2OVSOI}
\end{figure}
\begin{figure}
\includegraphics[scale=0.50,trim = 3mm 4mm 2mm 3mm,clip]{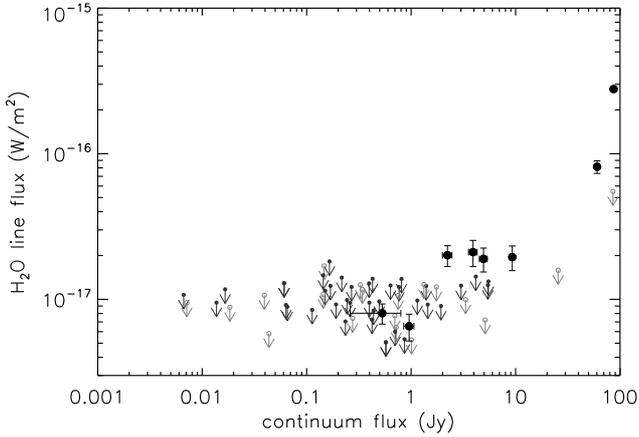}
  \caption{Plot of the 63.32 $\rm \mu m $ ortho-$\rm H_{2}O$ line flux versus 63$\mu$m continuum flux. Symbols as in Fig. \ref{H2OVSOI}.}
  \label{H2OLineFluxVScont}
\end{figure}
\begin{figure}[h]
    \includegraphics[scale=0.50,trim = 3mm 4mm 2mm 3mm,clip]{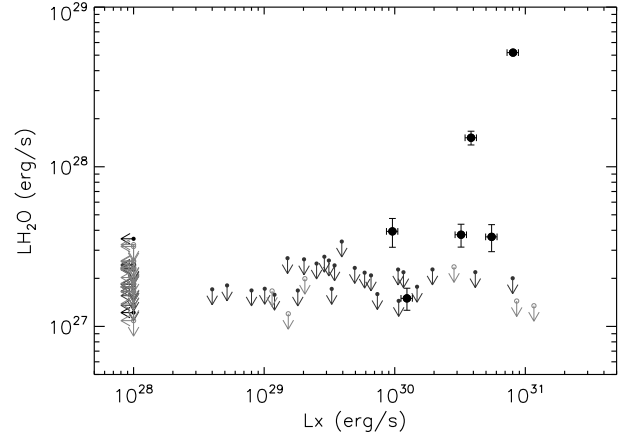}
  \caption{$\rm H_{2}O$  luminosity versus X-ray luminosity. Symbols as in Fig. \ref{H2OVSOI}.}
  \label{H2OLumVSLx}
\end{figure}
\begin{table}[th]
\caption{Probabilities for correlations between o-$\rm H_{2}O$ line intensity and stellar/disc parameters.}             
\label{CorrTab}      
\centering          
\begin{tabular}{llll}     
\hline\hline       
Observable & Points n$¼$ & Spearman's prob. &  Kendall's prob.\\ 
\hline                
$\rm L_{[OI]}$ & 68 & 0.0009 (0.4090 ) & 0.0000 (0.196) \\
63 $\rm \mu m$ flux & 64 & 0.0147 (0.2635 ) & 0.0002 (0.4567)\\ 
850 $\rm \mu m$ flux $\rm ^{(2)}$ & 57 & 0.0145 (0.2596) & 0.0131 (0.8496) \\
$\rm L_{star}$ $\rm ^{(3)}$ & 65& 0.1789 (0.1130) & 0.1980 (0.5535) \\
$\rm L_{X}$ $\rm ^{(3)}$ & 65 & 0.0225 (0.9912) & 0.0087 (0.6012) \\
$\rm \alpha (2 \mu m - 8 \mu m)$  $\rm ^{(4)}$ & 62 & 0.026 (0.1547) & 0.0023 (0.3850)\\ 
$\rm L_{[OI] 6300 \AA}$ & 27 & 0.1291  (0.3255) & 0.1008 (0.3450)  \\
\hline                  
\end{tabular}
\tablefoot{The values obtained for random populations are shown in brackets. Accurate only if N $>$ 30. (2):  850 $\rm \mu m $ continuum fluxes from \cite{Andrews2005}.  (3): Values from \cite{Gudel2007}. (4): SED slope from \cite{Luhman2010}.}
\end{table}
The o-$\rm H_{2}O$ emission is only present in spectra with
[\ion{O}{I}] detections. The targets \object{FS Tau}, \object{HL Tau}, and
\object{T Tau} display extended emission in the [\ion{O}{I}] 63.18 $\rm
\mu m$ line, but only T Tau show hints of extended emission in the $\rm 63.32 \mu m$
o-$\rm H_{2}O$ line. T Tau is an exceptional object.  It is a triple star system that drives at least two jets. 
It was the most line rich PMS star observed by ISO \citep{Lorenzetti2005}  and resembles more a hot core than a PMS star, 
since the continuum emission is quite extended ($\rm \sim$ 3.5 \arcsec at 70 $\rm \mu m$).  It is impossible to tell how much of the line emission comes from the discs,
 from the outflows, or even from the surrounding envelope.

To help us understand the origin of the o-$\rm H_{2}O$ emission, we
compared the line intensity with several star and disc (jet)
parameters. We computed survival analysis ranked statistics using the
ASURV code \citep{Feigelson1985,Isobe1986}. The result of this
analysis is summarised in Table \ref{CorrTab}. We also created random
populations to test the validity of the results.

The survival analysis shows a correlation between the o-$\rm H_{2}O$
line fluxes and the [\ion{O}{I}] line fluxes at a significance level
of 0.99 (see Fig. \ref{H2OVSOI}). This relationship suggests that both
lines have a similar origin. The $\rm H_{2}O$ emission is correlated
with the continuum emission at 63 $\rm \mu m$, but at a significance
level of 0.95 in the Spearman statistics
(Fig. \ref{H2OLineFluxVScont}). The 850 $\rm \mu m$ continuum flux can
be used as a proxy for the amount of dust present in the
disc. A survival analysis shows a possible correlation with the 850 $\rm
\mu m$ continuum flux, at a significance level of 0.95 in both
Spearman and Kendall statistics, although a large scatter is present.
A correlation with neither the stellar luminosity nor the spectral type
is found. There seems to be a weak correlation with the slope of the
SED measured between 2 $\rm \mu m$ and 8 $\rm \mu m$, used as a proxy for the
presence of hot dust. The significance of this correlation is dominated
by \object{T Tau}, which has the highest
o-$\rm H_{2}O$ flux in the sample. There is likely no link between mass loss
rate and $\rm L_{H_{2}O}$. The [\ion{O}{I}] luminosity at 6300 $\rm
\dot{A}$ is proportional to the mass loss rate \citep{Hartigan1995}.
Although the sample is too small to test this relationship conclusively, we note
that while $\rm L_{[OI]6300 \dot{A}}$ spans four orders of magnitude, the
$\rm H_{2}O$ luminosity spans only one order of magnitude.  
  Finally, the survival analysis statistics points to a possible
  relationship with the X-ray luminosity $L_x$
  (Fig. \ref{H2OLumVSLx}). While the Spearman probability for the real
  sample is only one order of magnitude smaller than for the random sample, the
  Kendall probability is two orders of magnitude smaller.
  Inspection of Fig.\ \ref{H2OLumVSLx} also shows that the o-$\rm H_{2}O$
  line flux is detected only for sources with X-ray luminosities
  higher than 10$^{30}$ erg s$^{-1}$, which is consistent with photochemical
  disc models that show that far-IR line fluxes increase
  significantly above this X-ray luminosity threshold
  \citep{Aresu2011}. We note that $\log L_x$ = 10$^{30}$ erg s$^{-1}$
  is above the median (mean) X-ray luminosity in Taurus ($\log L_x$ =
  29.8 (29.75) erg s$^{-1}$ respectively,
  \cite{Gudel2007}). Interestingly, more than half of the sources with 
  $L_x > 10^{30}$ erg s$^{-1}$ do not display the o-$\rm H_{2}O$ line.  This
  behaviour may stem from either (1) the different shape of the X-ray
  spectrum (hardness ratio), (2) the duty cycle of the flares
  responsible for the high levels of X-ray fluxes, and/or (3) that X-rays are not the only driver of $\rm H_{2}O$ chemistry and
  excitation, let alone any inner disc geometry and radiative transfer
  considerations. We caution that the correlation 
  with X-rays is significantly weaker when \object{TTau} is removed from the analysis.

All the stars detected in o-$\rm H_{2}O$ are outflow/jet sources, although three of them AA Tau, DL Tau,
  and RY Tau do not show any excess in [OI], i.e. all the emission is consistent with coming from 
  the disc (Howard et al., in prep). Two of these (AA Tau and DL Tau) are classified as outflow
   sources based on blue-shifted forbidden optical emission lines, although the emission is only slightly blue-shifted for AA Tau.
However, in the sample there are also some prominent
jet sources ($L_{\rm [OI]6300 \dot{A}} > 10^{-2}$ L$_{\sun}$) that show 
no hint of emission from this o-$\rm H_{2}O$ line, such as \object{DG Tau}, where
other $\rm H_{2}O$ lines have been detected. However, these $\rm H_{2}O$ lines are believed to originate in the outflow. 
Furthermore, the [OI] at 63.18 $\rm \mu m$ line is sometimes extended (Podio et al. in prep, Herczeg et al. 2011), while we find the o-$\rm H_{2}O$
to be unresolved, suggesting a more compact origin for the o-$\rm H_{2}O$ line.

We therefore assume that the o-$\rm H_{2}O$ emission at 63.32 $\rm \mu m$ originates from the
disc and ask whether it comes from the same gas reservoir as the hot $\rm H_{2}O$
lines observed by Spitzer. \citet{CarrNajita2011} detected six
out of eleven stars: \object{AA Tau}, \object{BP Tau}, \object{DK
  Tau}, \object{GI Tau}, \object{RW Aur}, and \object{UY Aur}. Our
sample contains all of their detected sources. We detected the 63.32 $\rm \mu m$ $\rm
H_{2}O$ emission only in \object{AA Tau} and \object{UY
  Aur}. \cite{Pontoppidan2010_0} reported $\rm H_{2}O$ detections toward three out of
eight stars in their sample: \object{DR Tau}, \object{AA Tau}, and
\object{IQ Tau}. We did not detect o-$\rm H_{2}$O in \object{IQ
  Tau}. \cite{Salyk2008}, \cite{Pontoppidan2009} and \cite{Meijerink2009} argued that
the \textit{Spitzer} hot $\rm H_{2}O$ emission comes from the
0.1 to 1.0 AU annular region of the disc. 

Assuming the same temperature, column density, and
emitting areas as \cite{CarrNajita2011}, i.e., 1AU,  we computed the
63.32$\mu$m LTE line flux in \object{AA Tau} and \object{UY
  Aur}. The model line emission is too optically thick to derive the $\rm H_{2}O$ mass
reliably. The size of the emitting region is instead estimated. In both cases, the 
model flux is ten times lower than measured. To recover the measured fluxes, the 
radius of the emitting area has to be about three times larger than the
value they quote, i. e., 3.0 AU for AA Tau and 3.5 AU for \object{UY Aur}, a result that 
is consistent with the lower excitation temperature of the line we observed. 

A radiation thermo-chemical model of a typical T~Tauri disc
  obtained with the {\sc ProDiMo} code
  \citep{Woitke2009B,Kamp2010,Thi2010,Aresu2011} predicts that the
  emission region of the 63.32 $\rm \mu$m o-H$_2$O line is of the order of
  3 AU, about five times larger than the Spitzer emission-line region
  (see Fig.~\ref{H2OProDiMo}).  One particular o-H$_2$O line at 15.738
  $\rm \mu$m was selected as a representative Spitzer mid-IR $\rm H_{2}O$
  emission line. This model does not intend to fit any particular object. A large non-LTE $\rm H_{2}O$
  ro-vibrational model calculation was included to consistently
  calculate the Spitzer as well as the Herschel $\rm H_{2}O$ emission lines
  by applying escape probability theory
  \citep{Faure2004,Faure2007,Faure2008}. At $\sim$~3~AU, the gas
  densities are high enough to excite the 63.32 $\rm \mu $m o-$\rm H_{2}O$
  line and the dust temperature close to the mid-plane is low enough
  for $\rm H_{2}O$ to freeze onto grains.

The 63.32 $\rm \mu m$ line could provide the missing link between the
mid-IR Spitzer detections of hot $\rm H_{2}O$ vapour in T Tauri discs
and the cold far-IR $\rm H_{2}O$ lines observed with
\textit{Herschel}. Searches for the lower excitation $\rm H_{2}O$
lines have proven to be less successful. To date, the only clear
detection of cold $\rm H_{2}O$ from a disc is \object{TW Hya} using
\textit{HIFI} \citep{Hogerheijde2011} , and Kamp et al. (in prep.),
using \textit{PACS}. Possible reasons for the lower
detection rate of cold $\rm H_{2}O$ might be the much lower $\rm H_{2}O$
abundances in the outer disc caused by the freeze-out of $\rm H_{2}O$ and/or
significant (vertical) settling of icy grains \citep{Bergin2010}. Observing 
the same molecule in transitions with very different excitation temperatures 
may trace it through a broader range of different radial zones in
protoplanetary discs.

\section{Conclusions}
We have detected o-$\rm H_{2}O$ emission at 63.32 $\rm \mu m$ in 8
T Tauri stars in a sub-sample of 68 stars located in Taurus. The detection rate is 
$\sim$24\% in the sub-sample with gas-rich discs. The $\rm H_{2}O$ emission appears to be
correlated with the continuum luminosities, the [\ion{O}{I}]  63.18 $\rm \mu m$ line 
fluxes, and the X-ray luminosities. The gas temperature (500-600K) and density 
needed to excite the observed o-$\rm H_{2}O$ line suggest that the line is coming from the inner parts of the discs
and from the upper layers of its atmosphere, where the disc is directly illuminated.
The correlation with X-rays flux and the role of X-ray emission in heating the gas, in particular during flares, needs to be investigated 
further. The typical size of the emitting region is estimated to be $ r \sim$ 3 AU, which is consistent with the typical location 
of the snow line in these objects. More effort is needed to detect several $\rm H_{2}O$ lines simultaneously in more objects to understand the full radial distribution of $\rm H_{2}O$ vapour in planet-forming discs.
\begin{acknowledgements}
We thank the anonymous referee for a constructive report that helped to improve the paper.
  This research has been funded by Spanish grants AYA 2010-21161-C02-02, CDS2006-00070 and
  PRICIT-S2009/ESP-1496. In France we thank ANR (contracts ANR-07-BLAN-0221 and ANR-2010-JCJC-0504-01); CNES; PNPS of CNRS/INSU. The Millennium Science Initiative (ICM) of the Chilean ministry of Economy (Nucleus P10-022-F) and an EC-FP7  grant (PERG06-GA-2009-256513) are also acknowledged.
\end{acknowledgements}

\bibliographystyle{aa} 
\bibliography{biblio.bib}

\begin{appendix}

\section{Non detections}
GASPS is a flux-limited survey. All sources have detection limits for the o-H$_2$O line at 63.32 $\mu$m around 
$10^{-17}$  W/$\rm m^{2}$, with an uncertainty of a factor of two both ways. The list of non detections is given in table \ref{NDTable}.

\begin{table}[h]
\caption{Source list, spectral types, and 63.32 $\rm \mu m$ o-$\rm H_{2}O$ line fluxes.}             
\label{NDTable}      
\centering          
\begin{tabular}{llllll}     
\hline\hline       
Name & SP. Type & $\rm o-H_{2}O$ flux     \\ 
\hline
-- & --  & $10^{-17} $ W/$\rm m^{2}$ \\
\hline                    
Anon 1 & M0 & $<$1.17 \\     
BP Tau & K7 & $<$0.94 \\
CIDA2 & M5.5 & $<$0.89 \\
CI Tau &  K7 & $<$0.92 \\
CoKu Tau/4 & K3 & $<$1.37 \\
CW Tau & K3  & $<$1.26 \\
CX Tau & M2.5 & $<$0.94 \\
CY Tau & M1.5 & $<$1.41 \\
DE Tau & M1 & $<$1.23 \\
DF Tau & M2 & $<$1.28 \\
DG Tau & K6? & $<$1.32 \\
DG Tau B & $<$ K6 & $<$1.25 \\
DH Tau & M1 & $<$1.07 \\
DK Tau & K6 & $<$0.53 \\
DL Tau & K7 & $<$1.14 \\
DM Tau & M1 & $<$0.89 \\
DN Tau & M0 & $<$0.77 \\
DO Tau & M0 & $<$1.21 \\
DP Tau & M0.5 & $<$0.92 \\
DQ Tau & M0 & $<$0.98 \\
DS Tau & K5 & $<$0.58 \\
FF Tau & K7 & $<$1.11 \\
FM Tau & M0 & $<$1.24 \\
FO Tau & M3.5 & $<$0.96 \\
FQ Tau & M3 & $<$0.91 \\
FT Tau & -- & $<$1.25 \\
FW Tau & M5.5  & $<$1.08 \\
FX Tau & M1 & $<$1.82 \\
GG Tau & M5.5 & $<$1.24 \\
GH Tau & M2 & $<$0.84 \\
GIK Tau & K7 & $<$0.74 \\
GM Aur & K7 & $<$1.21 \\
GO Tau & M0 & $<$1.23 \\
Haro 6-13 & M0 & $<$1.43 \\
Haro6-37& M1 &  $<$1.84 \\
HBC 347 & K1$^{1}$ & $<$1.69 \\
HBC 356 & K2$^{1}$ & $<$0.94 \\
HBC 358 &M3.5 & $<$1.29 \\
HK Tau & M0.5 & $<$0.90 \\
HN Tau & K5 & $<$0.64 \\
HO Tau & M0.5 & $<$1.46 \\
HV Tau & M1  & $<$0.72 \\
IP Tau & M0 & $<$0.83 \\
IQ Tau & M0.5 & $<$1.38 \\
04158+2805 & M5.25 & $<$1.06 \\
LkCa 1 & M4 & $<$1.06 \\
LkCa 3 & M1 & $<$0.94 \\
LkCa 4 & K7 & $<$0.98 \\
LkCa 5 & M2 & $<$1.16 \\
LkCa 7 & M0 & $<$0.88 \\
LkCa 15 & K5 & $<$0.52 \\
SAO 76428 & F8? & $<$1.58 \\
SU Aur & G2 & $<$0.72 \\
UX Tau & K5 & $<$0.99 \\
UZ Tau & M2 & $<$0.60 \\
V710 Tau & M0.5 & $<$1.17 \\
V773 Tau & K3 & $<$0.77 \\
V819 Tau & K7 & $<$1.21 \\
V927 Tau & M4.75 & $<$0.88 \\
VY Tau & M0 & $<$1.30 \\
\hline                  
\end{tabular}
\tablefoot{Spectral types from the compilation by \cite{Luhman2010}. (1): from \cite{Herbig1988}}
\end{table}

\section{Location of emission for $\rm H_{2}O$ lines }

\begin{figure}[hb]
    \includegraphics[width=9cm]{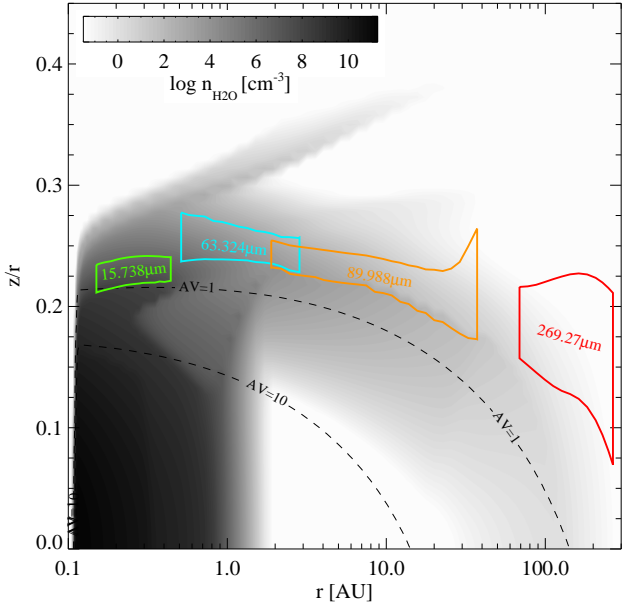}
    \caption{Location of the $\rm H_{2}O$ emission for a typical disc model
      ($M_*$=0.8M$_\odot$, $L_*$=0.7~L$_\odot$, $T_{\rm eff}$=4400K,
      UV excess $f_{\rm UV}$=0.01, X-ray luminosity $L_X$=10$^{30}$
      erg s$^{-1}$, $R_{\rm in}$=0.1AU, $R_{\rm out}$=300~AU, $M_{\rm
        disc}$=0.01~M$_\odot$, dust/gas=0.01, $\epsilon$=-1, scale
      height $H_0$=1~AU at $r_0$=10~AU, flaring $\beta=1.1$,
      astronomical silicate with uniform dust size distribution
      $a_{\rm min}$=0.05~$\mu$m, $a_{\rm max}$=1~mm, and
      index=3.5.). The 15.738 $\mu$m line is used as a typical $\rm H_{2}O$
      line detected by \textit{Spitzer}. We have marked the radii
      where the cumulative line flux reaches 15\% and 85\% with
      vertical lines, for the four selected $\rm H_{2}O$ lines. In addition, the
      horizontal lines mark the heights above the midplane where 15\%
      and 85\% of the line flux, from every vertical column, originate
      in. The encircled regions are hence responsible for 70\% x 70\%
      = 49\% of the total line flux (for a pole-on disc).}
  \label{H2OProDiMo}
\end{figure}

\end{appendix}

\end{document}